\documentclass[reprint,amsmath,amssymb,aps,pre,longbibliography]{revtex4-1}

\usepackage{graphicx}
\usepackage{dcolumn}
\usepackage{bm}
\usepackage{amsmath}
\usepackage{color}
\usepackage[utf8]{inputenc}
\usepackage{cancel}
\usepackage{easyReview}

\newcommand{\cvT}{\frac{\partial c_\rho}{\partial T}}
\newcommand{\cvr}{\frac{\partial c_\rho}{\partial \rho}}
\newcommand{\cvrr}{\frac{\partial (c_\rho\rho)}{\partial \rho}}

\begin{document}

\title{Transport coefficients from equilibrium thermodynamics: a unified formulation tested on the Lennard-Jones fluid}

\author{M. Hoyuelos}
\email{hoyuelos@mdp.edu.ar}

\affiliation{Instituto de Investigaciones F\'isicas de Mar del Plata (IFIMAR-CONICET), Departamento de F\'isica, Facultad de Ciencias Exactas y Naturales, Universidad Nacional de Mar del Plata, Funes 3350, 7600 Mar del Plata, Argentina}

\begin{abstract}
A thermodynamic transport theory (TTT) is applied, within a unified formulation, to the thermal conductivity, the shear viscosity, and the self-diffusion coefficient of simple fluids. In TTT, the Onsager coefficient of an interacting system is related to that of an ideal reference system through the ratio of the determinants of the Hessians of a thermodynamic potential. Each process is described by an effective transported density, the product of an ideal-form density, proportional to $\rho T^a$, and a factor that accounts for the departure of the exchanged amounts from their ideal values. The exponent $a$ is fixed by the degree of homogeneity of the transported quantity in the momentum for energy and momentum, and by the Lewis number of the dilute gas for composition. The departure factors are obtained from the thermal and mechanical parts of the enthalpy fluctuations of a fluid element; the mechanical part involves the thermal pressure, and, for self-diffusion, it leads to a factor $e^{s'_\text{ex}}$, where $s'_\text{ex}$ is the excess entropy per particle. Each transport coefficient is then the product of its dilute-gas value and a function of equilibrium thermodynamic properties, with no adjustable parameters. The predictions for the viscosity and the self-diffusion coefficient of the Lennard-Jones fluid are compared with molecular dynamics results at supercritical temperatures and with the revised Enskog theory (RET) for Mie fluids. The accuracy of TTT is similar to that of RET: it is better at intermediate temperatures for the viscosity and at high temperatures for self-diffusion, and worse close to the critical point, where the theory is very sensitive to the curvature of the free energy.
\end{abstract}


\maketitle

\section{Introduction}

The prediction of the transport coefficients of dense fluids for different interaction potentials is a long-standing problem of statistical mechanics. In the dilute-gas limit, it is solved by the Chapman-Enskog theory, based on the Boltzmann equation, which gives the thermal conductivity, the viscosity, and the self-diffusion coefficient in terms of collision integrals \cite{chapman,Hirschfelder}. In this limit, the range of the interaction is small compared with the mean free path, and momentum and energy are carried by the particles in free flight. At higher densities, the transfer of momentum and energy by the interparticle forces, over the distance between interacting particles, becomes comparable to that by free flight, and correlations between particles play an essential role \cite{hansen2,dorfmanbook}. The theory of Enskog for hard spheres (HS) \cite{enskog,chapman} introduces the effect of density through the radial distribution function at contact and the collisional transfer of momentum and energy across the particle diameter. The revised Enskog theory (RET) \cite{Beijeren2} extends it consistently to mixtures, and Jervell and Wilhelmsen developed a RET for Mie fluids, in which the radial distribution function at contact is obtained from the perturbation theory underlying the SAFT-VR Mie equation of state (EOS) \cite{jervell,lafitte}; in a more recent version, the collisional contributions are evaluated with transfer lengths for momentum and energy obtained from the dynamics of binary collisions \cite{jervell2024predicting}. Other routes require adjustable parameters; some examples are empirical correlations of simulation data \cite{bugel,galliero,chaparro} and excess-entropy scaling \cite{rosenfeld,rosenfeld3,dyre}, in which the reduced transport coefficients are expressed as functions of the excess entropy.

A different approach is based on equilibrium thermodynamics. The idea that the effect of the interactions on transport coefficients can be largely described by thermodynamic factors goes back to the work of Darken on diffusion in alloys \cite{darken}. In recent years, several transport coefficients of simple fluids were expressed as the product of their ideal value and a function of the thermodynamic state: self-diffusion \cite{marchioni,sampayo}, viscosity \cite{marchioni2}, and the transport coefficients of the HS fluid \cite{hoyuelosHS}. A general framework for this kind of description was developed by Di Muro and Hoyuelos \cite{dimuro-hoyuelos}: the Onsager matrix of an interacting system is equal to that of an ideal reference system, defined by the same transported variables, multiplied by the ratio of the determinants of the Hessians of the relevant thermodynamic potential in the ideal and real systems. This framework is referred to as the thermodynamic transport theory (TTT). It was recently applied to the thermal conductivity of the Lennard-Jones (LJ) fluid, with an accuracy similar to that of the most recent version of RET \cite{alesLJ}.

The application of TTT to a fluid in continuous space requires the identification of the transported variables, which must be the same functions of the state in the real and ideal systems. In this paper, a unified formulation for the thermal conductivity, the shear viscosity, and the self-diffusion coefficient of simple fluids is presented, in which each process is described by an effective transported density. It is the product of an ideal-form density, which fulfills the condition of the theory, and a departure factor, which accounts for the difference between the amounts exchanged in the real and ideal systems. For the thermal conductivity, the result of Ref.~\cite{alesLJ} is recovered, and, for HS, the results of Ref.~\cite{hoyuelosHS}.

The predictions for the viscosity and the self-diffusion coefficient are compared with molecular dynamics (MD) results for the LJ fluid at supercritical temperatures, and with those of RET. The paper is organized as follows. In Sec.~\ref{s.ttt}, we review the TTT and its application to a fluid, where the number of particles of a cell is not a transported variable. The unified formulation is presented in Sec.~\ref{s.unified}, and the equilibrium fluctuations of a fluid element, from which the departure factors are obtained, in Sec.~\ref{s.fluct}. The thermal conductivity and the viscosity are considered in Sec.~\ref{s.lambdaeta}, and the self-diffusion coefficient in Sec.~\ref{s.D}. The comparison with numerical results is presented in Sec.~\ref{s.comp}. The results and the hypotheses on which they rest are summarized in Sec.~\ref{s.summary}, and the conclusions are given in Sec.~\ref{s.conclusions}.

\section{Review of thermodynamic transport theory}
\label{s.ttt}

In this section we summarize the thermodynamic transport theory (TTT) introduced in Ref.~\cite{dimuro-hoyuelos}, which is the starting point of the present work. The system is divided into cells, each in local thermodynamic equilibrium, that exchange extensive quantities with their neighbors; for a given cell, the rest of the system acts as a reservoir that fixes the intensive parameters. The state of a cell is specified by the extensive variables $\vec X=(X_1,\dots,X_n)$ that can be exchanged between neighboring cells, and by the conjugate entropic intensive parameters $\vec Y=\partial S/\partial \vec X$, so that
\begin{equation}
dS = \vec Y\cdot d\vec X,
\label{e.dS}
\end{equation}
where $S$ is the entropy of the cell. In the linear regime, the fluxes of the transported quantities are proportional to the thermodynamic forces, $\vec F=-\nabla \vec Y$,
\begin{equation}
\vec J = -L\cdot \vec F = L\cdot \nabla \vec Y,
\label{e.onsager}
\end{equation}
where $L$ is the (symmetric, positive definite) Onsager matrix. The entropy Hessian,
\begin{equation}
H = \frac{\partial^2 S}{\partial \vec X^2} = \frac{\partial \vec Y}{\partial \vec X},
\label{e.hessian}
\end{equation}
is negative definite in a thermodynamically stable state and determines the equilibrium fluctuations of $\vec X$ in a cell, whose covariance matrix is $C=-H^{-1}$ (with $k_B=1$) \cite{callen,reichl}.

The central result of Ref.~\cite{dimuro-hoyuelos} relates $L$ to the Onsager matrix $L_\text{id}$ of an ideal reference system, in which interactions are neglected. It is obtained by imposing detailed balance on the elementary events in which a small amount $\vec\delta$ of the extensive quantities is transferred between two neighboring cells. The probabilities of the initial and final macrostates are computed with the statistical-mechanical entropy, which contains, besides the extensive thermodynamic entropy, a nonextensive term of order $\ln N$ \cite{hoyuelos}. Expanding up to order $N^{-1}$ and using the ideal reference system to eliminate an unknown function of the state, one finds that the equilibrium rate of the exchange events is
\begin{equation}
\bar W_{\vec\delta} = \bar W^\text{id}_{\vec\delta}\, \frac{\det C}{\det C_\text{id}},
\label{e.rate}
\end{equation}
where the factor $\det C/\det C_\text{id} = \det H_\text{id}/\det H$ comes from the excess part of the nonextensive term of the entropy. The Onsager matrix of the ideal system, $L_\text{id}$, is built from the rates $\bar W^\text{id}_{\vec\delta}$ and the amounts $\vec\delta$, and that of the interacting system is
\begin{equation}
L = \frac{\det H_\text{id}}{\det H}\, L_\text{id}.
\label{e.main}
\end{equation}
The effect of the interactions on the Onsager matrix thus reduces to a scalar factor that depends only on equilibrium properties, the ratio of the volumes of the fluctuation ellipsoids in the real and ideal systems, while all the dynamical information is contained in $L_\text{id}$. Equation~\eqref{e.main} reproduces the Darken equation for substitutional diffusion in solids and the nondiagonal diffusion matrix of the Zhdanov model for surface diffusion \cite{dimuro-hoyuelos}.

The elimination of this unknown function deserves a comment, since it is the step that has to be reconsidered in a fluid. The equilibrium rate has the form \cite{dimuro-hoyuelos}
\begin{equation}
\bar W_{\vec\delta} = e^{-2\Delta S + f},
\label{e.Wf}
\end{equation}
where $\Delta S$ is the nonextensive term of the entropy of a cell, with $e^{-2\Delta S}\propto\det C$ \cite{hoyuelos}, and $f$ is a function that is not fixed by detailed balance; it depends on the state only through $\vec X$. In Ref.~\cite{dimuro-hoyuelos}, the ideal reference system is the system with the same variables $\vec X$ and the same function $f$, $f_\text{id}(\vec X)=f(\vec X)$, for the same state and the same transferred amount $\vec\delta$. This is the case when the interactions modify the rates only through the entropy, so that the whole excess of the rate with respect to the ideal system is contained in the nonextensive term $\Delta S$, as in the lattice dynamics considered there. Equation~\eqref{e.rate} then follows. The equality $f=f_\text{id}$ at finite density is an assumption, which is reconsidered below for a fluid.

%
Since the rates of the real and ideal systems are compared in the same state, the transported variables of the ideal system must be identical, as functions of the state, to those of the real system: 
\begin{equation}\label{e.condition}
	\vec X_\text{id}=\vec X.
\end{equation}
%

\subsection{Non-transported variables}
\label{s.nontransp}

The set $\vec X$ in Eq.~\eqref{e.main} contains the quantities that are transported by exchange between cells. An extensive variable $X'$ that is not transported, with conjugate parameter $Y'$, can be found in two different situations.

In the first situation, the one considered in Ref.~\cite{dimuro-hoyuelos}, the parameter $Y'$ is fixed by the reservoir and $X'$ plays no further role; an example is the energy in isothermal diffusion on a lattice. The probability of the transported variables is then determined by the Massieu function $\Phi=S-X'Y'$ \cite{hoyuelos}, and Eq.~\eqref{e.main} holds with $H$ the Hessian of $\Phi$ with respect to the transported variables only. The dimension of the Hessian is reduced, and $C=-H^{-1}$ is the covariance matrix of the transported variables.

In the second situation, $X'$ has no diffusive flux, and therefore no Onsager coefficient, but it is nevertheless exchanged between cells by a mechanism whose equilibrium rate is known. This is the case of the number of particles $N$ in a cell of fixed volume of a one-component fluid: the particle flux has no dissipative part (see Sec.~\ref{s.cells}), but particles continually cross the boundaries of the cell. We consider, for simplicity, a single transported variable $X$, with $dS=Y_X\,dX+Y_N\,dN$. A cell exchanges $X$ and $N$ with its neighbors through two channels: the events that transfer an amount $\delta$ of $X$ at fixed $N$, and the crossings of the boundary, which change $N$ by one unit. For each channel, the other variable is in contact with the reservoir, as in the first situation, and Eq.~\eqref{e.Wf} gives
\begin{equation}
\frac{\bar W_X}{\bar W_X^\text{id}} = \frac{C_{XX}}{C_{XX}^\text{id}}\, e^{f_X-f_X^\text{id}},
\qquad
\frac{\bar W_N}{\bar W_N^\text{id}} = \frac{C_{NN}}{C_{NN}^\text{id}}\, e^{f_N-f_N^\text{id}},
\label{e.channels}
\end{equation}
where $C_{XX}=-(\partial X/\partial Y_X)_{Y_N}$ and $C_{NN}=-(\partial N/\partial Y_N)_{Y_X}$ are the variances of $X$ and $N$ in a cell.

The rate of the second channel is known exactly. In a classical fluid in equilibrium, the momenta of the particles have the Maxwell distribution and are independent of the positions, for any interaction potential; this is a consequence of the separation of the energy into a kinetic part, which depends only on the momenta, and a potential part, which depends only on the positions. The number of particles that cross a fixed surface in one direction, per unit area and unit time, is therefore the same in the fluid and in the ideal gas at the same density and temperature, so that $\bar W_N=\bar W_N^\text{id}$ and
\begin{equation}
e^{f_N-f_N^\text{id}} = \frac{C_{NN}^\text{id}}{C_{NN}}.
\label{e.fN}
\end{equation}
The function $f$ of a fluid is thus modified by the interactions, by an amount that vanishes in the dilute limit. This is the difference with the lattice models of Ref.~\cite{dimuro-hoyuelos}, where $f=f_\text{id}$ and the jump rate of a particle is reduced by the factor $C_{NN}/C_{NN}^\text{id}$.

The following assumption is introduced: the modification of the function $f$ is the same for the two channels,
\begin{equation}
f_X-f_X^\text{id} = f_N-f_N^\text{id},
\label{e.hyp}
\end{equation}
i.e., it represents a change of the time scale of the exchanges between two cells, which does not depend on the quantity exchanged. The condition of Ref.~\cite{dimuro-hoyuelos}, $f=f_\text{id}$ for every channel, is a particular case of Eq.~\eqref{e.hyp}. From Eqs.~\eqref{e.channels}--\eqref{e.hyp},
\begin{equation}
\bar W_X = \bar W_X^\text{id}\; \frac{C_{XX}/C_{NN}}{C_{XX}^\text{id}/C_{NN}^\text{id}}.
\label{e.rate2}
\end{equation}
The ratio of variances can be written in terms of the Massieu function $\Phi=S-NY_N$, whose natural variables are $(X,Y_N)$. From $d\Phi=Y_X\,dX-N\,dY_N$, the Hessian of $\Phi$ with respect to both natural variables, $H_\Phi$, has the determinant
\begin{equation}
\det H_\Phi = -\left(\frac{\partial Y_X}{\partial X}\right)_{Y_N}\left(\frac{\partial N}{\partial Y_N}\right)_{Y_X} = -\frac{C_{NN}}{C_{XX}}.
\label{e.detHPhi}
\end{equation}
Building the Onsager coefficient from the rates as before, we obtain
\begin{equation}
L = \frac{\det H_\Phi^\text{id}}{\det H_\Phi}\, L_\text{id},
\label{e.main2}
\end{equation}
which has the form of Eq.~\eqref{e.main} with the entropy Hessian replaced by the Hessian of the Massieu function with respect to all its natural variables; the dimension of the Hessian is not reduced. Unlike $H$, the matrix $H_\Phi$ is not negative definite, since $\Phi$ is concave in $X$ and convex in $Y_N$, and $-H_\Phi^{-1}$ is not a covariance matrix; the ratio of determinants in Eq.~\eqref{e.main2} is nevertheless positive. 

The hypothesis \eqref{e.hyp} is supported by the structure of this result. In thermodynamic geometry, the Hessian of the entropy defines a metric on the space of equilibrium states \cite{ruppeiner}, and a total Legendre transformation describes the same geometry in dual coordinates \cite{amari}. A partial Legendre transformation, such as the one from $S$ to $\Phi$, leads instead to the Hessian of the transformed potential with respect to its natural variables, which is indefinite and is not equivalent to the metric of the entropy representation; the different partial transforms correspond to different ensembles \cite{salamon,bravetti}. Equation~\eqref{e.main2} states that, with the hypothesis \eqref{e.hyp}, the Onsager coefficient of a fluid is determined by the geometry associated with the representation $(X,Y_N)$, in the same way as, on a lattice, it is determined by the potential of the variables that are not fixed by the reservoir \cite{dimuro-hoyuelos}.

The events of the first channel are exchanges of $X$ at fixed $N$, in which the crossings are compensated, as corresponds to the processes at uniform density considered in Sec.~\ref{s.unified}. Since $N$ has no diffusive flux, the events that contribute to $L$ have $\delta_N=0$, and the single crossings enter only through their equilibrium rate, used in Eq.~\eqref{e.fN}. Thus, Eq.~\eqref{e.fN} is exact and the separation into channels follows from the absence of a diffusive particle flux, while Eq.~\eqref{e.hyp} is a hypothesis. Equation~\eqref{e.main2} is the form of the theory that was used for HS in Ref.~\cite{hoyuelosHS} and for the thermal conductivity of the LJ fluid in Ref.~\cite{alesLJ}, and it is the one used in the present work.

\section{Unified formulation}
\label{s.unified}

In this section we apply the theory, in the form of Eq.~\eqref{e.main2}, to a simple fluid in continuous space and obtain a general expression for the ratio $\zeta/\zeta_\text{id}$, where $\zeta$ represents the thermal conductivity $\lambda$, the shear viscosity $\eta$, or the self-diffusion coefficient $D$, and $\zeta_\text{id}$ is the corresponding dilute-gas value. The formulation follows the one developed for HS in Ref.~\cite{hoyuelosHS}, generalized to fluids whose specific heat depends on the thermodynamic state and whose compressibility factor depends on temperature, as done for the thermal conductivity in Ref.~\cite{alesLJ}. The specific choices for each transport coefficient are presented in Secs.~\ref{s.fluct}--\ref{s.D}.

\subsection{Cells and thermodynamic variables}
\label{s.cells}

We consider a one-component simple fluid. The Gibbs relation for a small cell, written in terms of densities, is
\begin{equation}
ds = \frac{1}{T}\,du - \frac{\mu}{T}\,d\rho,
\label{e.gibbs}
\end{equation}
where $s=S/V$, $u=U/V$, and $\rho=N/V$ are the entropy, internal-energy, and particle-number densities, $T$ is the temperature, and $\mu$ is the chemical potential per particle. We use units with $k_B=1$.

In the lattice formulation of Ref.~\cite{dimuro-hoyuelos}, cells have fixed positions and exchange particles. As in Ref.~\cite{hoyuelosHS}, we consider cells of fixed volume, large enough for Eq.~\eqref{e.gibbs} to hold locally and small compared with the characteristic length of the perturbations, and we work in the comoving reference frame, in which the equilibrium mean velocity vanishes. In a one-component fluid, the particle flux is the momentum density divided by $m$; it vanishes on average in the comoving frame and has no dissipative part, so that the evolution of $\rho$ is completely determined by the continuity equation. The particle density is therefore not transported, in the sense that it has no Onsager coefficient. The number of particles in a cell nevertheless fluctuates, since particles cross its boundaries. This is the second situation of Sec.~\ref{s.nontransp}, and the density is treated with Eq.~\eqref{e.main2}.

To describe self-diffusion, we assign labels A and B to otherwise identical particles, with densities $\rho_A$ and $\rho_B$, and introduce $\rho=\rho_A+\rho_B$ and the composition density $\rho_D=\rho_A-\rho_B$. With $\mu_A=\mu+T\ln(\rho_A/\rho)$ and $\mu_B=\mu+T\ln(\rho_B/\rho)$, the Gibbs relation of the binary mixture is
\begin{equation}
ds = \frac{1}{T}\,du - \frac{\mu_N}{T}\,d\rho - \frac{\mu_D}{T}\,d\rho_D,
\end{equation}
with
\begin{equation}
\frac{\mu_N}{T} = \frac{\mu}{T} + \frac{1}{2}\ln\left[\frac{1-r^2}{4}\right], \qquad
\frac{\mu_D}{T} = \frac{1}{2}\ln\left(\frac{1+r}{1-r}\right),
\label{e.muD}
\end{equation}
where $\mu_N=(\mu_A+\mu_B)/2$, $\mu_D=(\mu_A-\mu_B)/2$, and $r=\rho_D/\rho$. The numbers of A and B particles in the whole system are equal, so that the equilibrium composition satisfies $r=0$ and $\mu_D=0$; on this equilibrium manifold, the Gibbs relation reduces to Eq.~\eqref{e.gibbs} with an irrelevant additive constant in $\mu/T$, while local composition fluctuations remain possible \cite{hoyuelosHS}.

We consider three transport processes, each driven at uniform particle density. Heat conduction is the transport of internal energy driven by a temperature gradient in a fluid at rest. Shear viscosity is the transport of transverse momentum density $g_\perp=\rho m v_\perp$, where $m$ is the particle mass and the velocity perturbation $v_\perp$ is perpendicular to the wave vector of the perturbation, so that, by the linearized continuity equation, it does not generate density variations. Self-diffusion is the transport of $\rho_D$ at uniform $\rho$. The corresponding transport coefficients are defined by
\begin{equation}
\vec J_u = -\lambda\nabla T, \qquad \vec J_{g} = -\eta \nabla v_\perp, \qquad \vec J_D = -D\nabla \rho_D,
\label{e.coefs}
\end{equation}
where $\vec J_g$ is the flux of transverse momentum along the direction of the wave vector.

As a first step, we use $(T,\rho)$ as thermodynamic coordinates. Using $du=\rho c_\rho\, dT + (\mu-T\mu_T)\,d\rho$, where $c_\rho$ is the specific heat per particle at constant density and $\mu_T=(\partial\mu/\partial T)_\rho$, Eq.~\eqref{e.gibbs} becomes
\begin{equation}
ds = \frac{\rho c_\rho}{T}\,dT - \mu_T\, d\rho.
\label{e.dsTrho}
\end{equation}
Here and in the following, partial derivatives with respect to $T$ and $\rho$ without subscripts are taken at constant $\rho$ and constant $T$, respectively.

\subsection{Effective transported densities}
\label{s.effective}

The densities $u$, $g_\perp$, and $\rho_D$ cannot be used directly as the transported variable of Sec.~\ref{s.ttt}. What the cells actually exchange are small amounts of these quantities, whose typical size is set by the spontaneous fluctuations; when a force, represented by a small gradient of an associated variable, is applied, the size of these amounts is still given by the fluctuations, whose connection with the transported quantities is discussed in Sec.\ \ref{s.fluct}. This distinction is relevant here, since TTT relates the transport coefficients to the second derivatives of a thermodynamic potential with respect to the transported variables, which depend on how these variables change with temperature and density. To obtain the correct thermodynamic factor, the variable used in the theory must describe the amounts actually exchanged, while its flux is, by construction, the same as the flux of the total density.

Each process is therefore described with an effective density, a function of the thermodynamic state that quantifies the portion of the transported quantity, per unit volume, that participates in the exchanges. It is constructed in two steps: an ideal-form density $\tilde x$, which fulfills Eq.~\eqref{e.condition} and to which the theory is applied, and a factor $R$ that accounts for the departure of the exchanged amounts from their ideal values.

\subsubsection{Ideal-form density}

The condition \eqref{e.condition} is fulfilled by quantities that depend only on the momenta of the particles. Since the momenta have the Maxwell distribution and are independent of the positions for any interaction potential (Sec.~\ref{s.nontransp}), the density of such a quantity is the same function of $(\rho,T)$ in the fluid and in the ideal gas. If the quantity carried by a particle is a homogeneous function of degree $2a$ of its momentum, its average is proportional to $(mT)^a$, and the ideal-form density is
\begin{equation}
\tilde x = \epsilon\, \rho\, T^a,
\label{e.xtilde}
\end{equation}
where $\epsilon$ does not depend on the thermodynamic state and cancels from the final results (in viscosity and self-diffusion, it varies slightly in space; see Sec.~\ref{s.diffusivity}). The only input is the degree of homogeneity: for heat conduction, the quantity is the kinetic energy $p^2/(2m)$, and $a=1$; for viscosity, the mean transverse momentum vanishes in equilibrium, and the relevant scale is that of the transverse momentum carried by the particles that cross a boundary, whatever its sign, which is proportional to $\rho\langle|p_\perp|\rangle\propto\rho\sqrt{mT}$, so that $a=1/2$. Thus,
\begin{equation}
\tilde x_u = \epsilon_u\,\rho T, \qquad \tilde x_\perp = \epsilon_\perp\, \rho\sqrt{mT}
\label{e.xue}
\end{equation}
(the factor $\sqrt m$ can be absorbed into $\epsilon_\perp$).

Composition behaves differently. The label of a particle does not depend on its momentum and the previous ideas do not apply. The ideal-form density for composition is taken as $\tilde x_D=\epsilon_D\,\rho T^{a_D(T_0)}$, where $T_0$ is the equilibrium temperature, around which the temperature $T$ of a cell fluctuates, and the exponent $a_D$ is determined in Sec.~\ref{s.aD} by the coupling between a diffusion event and the local thermal relaxation in the dilute gas. It depends on $T_0$ through the dilute-gas transport coefficients, but not on $T$, and it is therefore a constant in the Hessian.

\subsubsection{Departure factor}

We call $x$ the effective density of the physical quantity transferred in an exchange between cells. It differs from $\tilde{x}$ by a factor $R$:
\begin{equation}
	x = \tilde x\, R,
	\label{e.xR}
\end{equation}
and its flux is identified with the physical flux of the transported quantity,
\begin{equation}
	\vec J_{x_u}=\vec J_u, \qquad \vec J_{x_\perp}=\vec J_g, \qquad \vec J_{x_D}=\vec J_D.
	\label{e.fluxid}
\end{equation}
$R$ is a function of the state that represents the departure from ideality, with $R\to 1$ in the dilute-gas limit. It is obtained, from equilibrium fluctuations, for viscosity and thermal conductivity in Sec.~\ref{s.lambdaeta}. For composition, it follows from a balance between the chemical-potential difference of the labeled components and the mechanical part of the enthalpy fluctuations (Sec.~\ref{s.D}).

\subsection{Onsager coefficient of the ideal-form density}
\label{s.onsager}

To write the Gibbs relation in terms of $\tilde x$ and $\rho$, we use $d\tilde x = \epsilon a\rho T^{a-1}\,dT + \epsilon T^a\, d\rho$ to eliminate $dT$ from Eq.~\eqref{e.dsTrho}:
\begin{equation}
ds = \frac{c_\rho}{\epsilon a T^a}\, d\tilde x - \left(\mu_T + \frac{c_\rho}{a}\right) d\rho.
\label{e.dsx}
\end{equation}
The intensive parameter conjugate to $\tilde x$ is
\begin{equation}
\tilde Y = \left(\frac{\partial s}{\partial \tilde x}\right)_\rho = \frac{c_\rho}{\epsilon a T^a}.
\label{e.Ytilde}
\end{equation}
As discussed in Sec.~\ref{s.nontransp}, the density is a non-transported variable. The corresponding Massieu function is obtained with the Legendre transform
\begin{equation}
\phi = s + \rho\left(\mu_T + \frac{c_\rho}{a}\right),
\qquad
d\phi = \tilde Y\, d\tilde x + \rho\, d\left(\mu_T + \frac{c_\rho}{a}\right).
\label{e.phi}
\end{equation}
There is a single transported quantity, $\tilde x$, and the Onsager matrix reduces to a scalar, $L_{\tilde x}$. The Hessian of $\phi$ with respect to its natural variables, which enters Eq.~\eqref{e.main2}, is calculated in Appendix~\ref{a.hessian}. The result is conveniently expressed in terms of the specific heat per particle along the path of constant $\tilde x$, i.e., of constant $\rho T^a$,
\begin{equation}
c_{\tilde x} = T\left(\frac{\partial s'}{\partial T}\right)_{\tilde x} = c_\rho + a\,\frac{\partial (ZT)}{\partial T},
\label{e.ca}
\end{equation}
where $s'=s/\rho$ is the entropy per particle and $Z=P/(\rho T)$ is the compressibility factor. For $a=1$, $c_{\tilde x}$ is the specific heat at constant $\rho T$ introduced in Ref.~\cite{alesLJ}; in the ideal gas, $c_{\tilde x}^\text{id}=c_\rho^\text{id}+a$, with $c_\rho^\text{id}=3/2$ for a monatomic fluid. Using Eq.~\eqref{e.main2}, we obtain
\begin{equation}
L_{\tilde x} = \frac{c_\rho^\text{id}}{c_{\tilde x}^\text{id}}\;
\frac{a c_{\tilde x} - T\left(\frac{\partial c_{\tilde x}}{\partial T}\right)_{\tilde x}}{a c_\rho - T\cvT}\; L^\text{id}_{\tilde x}.
\label{e.Lx}
\end{equation}

\subsection{Diffusivity of the ideal-form density}
\label{s.diffusivity}

A flux requires a gradient of the intensive parameter which, from Eqs.~\eqref{e.xtilde} and \eqref{e.Ytilde}, can be written as
\begin{equation}
\tilde Y = \frac{c_\rho}{\epsilon a T^a} = \frac{\rho\, c_\rho}{a\,\tilde x}.
\label{e.Yx}
\end{equation}
In heat conduction, $\tilde x_u$ has a gradient because the temperature has one, at uniform density. In viscosity and self-diffusion, the density and the temperature are uniform, but there is an imbalance of transverse momentum or of composition, which does not change the thermodynamic state. In this case, the imbalance enters the effective density, $\tilde x=\epsilon\rho T^a$, through $\epsilon$: this factor varies slightly in space, generating the force that drives transverse momentum and composition currents.

\subsubsection{Heat conduction}

For heat conduction, $\tilde x_u=\epsilon_u\rho T$ ($a=1$), and its gradient is produced by a temperature gradient at uniform density. According to Eq.\ \eqref{e.onsager}, its flux is
\begin{equation}
\vec J_{\tilde x_u} = L_{\tilde x_u}\left(\frac{\partial \tilde Y_u}{\partial \tilde x_u}\right)_\rho \nabla\tilde x_u = -D_{\tilde x_u}\nabla\tilde x_u,
\label{e.Jxt}
\end{equation}
where $D_{\tilde x_u}$ is the diffusivity of $\tilde{x}_u$ and, from Eq.~\eqref{e.Yx},
\begin{equation}
\left(\frac{\partial \tilde Y_u}{\partial \tilde x_u}\right)_\rho = -\frac{c_\rho - T\cvT}{\epsilon_u^2 \rho\, T^{2}}.
\label{e.dYdx}
\end{equation}
In the ideal system, the density $\tilde x_u$ is the same, and $\tilde Y_u^\text{id}=c^\text{id}_\rho/(\epsilon_u T)$, with $c_\rho^\text{id}$ constant, so that $(\partial \tilde Y_u^\text{id}/\partial\tilde x_u)_\rho=-c_\rho^\text{id}/(\epsilon_u^2 \rho T^{2})$. The ratio of diffusivities in the real and ideal systems is
\begin{equation}
\frac{D_{\tilde x_u}}{D^\text{id}_{\tilde x_u}}
= \frac{L_{\tilde x_u}}{L^\text{id}_{\tilde x_u}}\,
\frac{c_\rho - T\cvT}{c^\text{id}_\rho}.
\label{e.DratioL}
\end{equation}
Using Eq.~\eqref{e.Lx} with $a=1$, the factor $c_\rho - T\,\partial c_\rho/\partial T$ cancels, and we obtain
\begin{equation}
\frac{D_{\tilde x_u}}{D^\text{id}_{\tilde x_u}} = \frac{1}{c_{\tilde x_u}^\text{id}}\left[c_{\tilde x_u} - T\left(\frac{\partial c_{\tilde x_u}}{\partial T}\right)_{\tilde x_u}\right],
\label{e.Dratio}
\end{equation}
with $c_{\tilde x_u}^\text{id}=c_\rho^\text{id}+1=5/2$ for a monatomic fluid. For practical calculations, it is convenient to write it in terms of derivatives at constant density and constant temperature (see Appendix~\ref{a.hessian}):
\begin{equation}
\frac{D_{\tilde x_u}}{D^\text{id}_{\tilde x_u}} = \frac{c_\rho - T\cvT + 2\rho\cvr + \Gamma_\text{th}}{c_\rho^\text{id}+1},
\label{e.Dratio2}
\end{equation}
where $\Gamma=\partial(\rho Z)/\partial\rho$ is the thermodynamic factor and $\Gamma_\text{th}=\partial(\Gamma T)/\partial T$; the meaning of $\Gamma_\text{th}$ is discussed in Sec.~\ref{s.fluct}. When the EOS is given in terms of the Helmholtz free energy, Eq.~\eqref{e.Dratio2} requires derivatives up to third order. Equation~\eqref{e.Dratio} coincides with the result of Ref.~\cite{alesLJ}.

\subsubsection{Viscosity and self-diffusion}

For viscosity and self-diffusion, $\tilde x$ is $\tilde x_\perp$, with $a=1/2$, or $\tilde x_D$, with $a=a_D$ (Sec.~\ref{s.aD}). Since the transport takes place at uniform density and temperature, the gradient of $\tilde x$ comes from the spatial variation of $\epsilon$, $\nabla\tilde x=\rho T^a\nabla\epsilon$. Since $T$ and $c_\rho$ are uniform, the intensive parameter of Eq.~\eqref{e.Yx} varies only through its explicit dependence on $\epsilon$:
\begin{equation}
\nabla\tilde Y = -\frac{c_\rho}{\epsilon^2 a T^a}\,\nabla\epsilon = -\frac{\tilde Y}{\tilde x}\,\nabla\tilde x .
\label{e.gradYeps}
\end{equation}
The flux of $\tilde x$ is then $\vec J_{\tilde x}=L_{\tilde x}\nabla\tilde Y=-D_{\tilde x}\nabla\tilde x$, with
\begin{equation}
D_{\tilde x} = L_{\tilde x}\,\frac{\tilde Y}{\tilde x} = L_{\tilde x}\,\frac{\rho\, c_\rho}{a\,\tilde x^2}.
\label{e.Deps}
\end{equation}
In the ideal system, the density $\tilde x$ is the same and $c_\rho^\text{id}$ replaces $c_\rho$. The ratio of diffusivities is
\begin{equation}
\frac{D_{\tilde x}}{D^\text{id}_{\tilde x}}
= \frac{L_{\tilde x}}{L^\text{id}_{\tilde x}}\,\frac{c_\rho}{c^\text{id}_\rho}
= \frac{c_\rho}{c_\rho^\text{id}+a}\;
\frac{a c_\rho - T\cvT + 2a\rho\cvr + a^2\Gamma_\text{th}}{a c_\rho - T\cvT},
\label{e.DratioU}
\end{equation}
where Eq.~\eqref{e.Lx} and Appendix~\ref{a.hessian} were used in the second equality. Equation~\eqref{e.DratioU} differs from Eq.~\eqref{e.Dratio2} because of the way in which $\tilde Y$ varies in space: in heat conduction, through the temperature, including the dependence of $c_\rho$ on it; in viscosity and self-diffusion, only through $\epsilon$.

\subsection{Transport coefficients}
\label{s.sigma}

We now connect the diffusivity of $\tilde x$ with the transport coefficients defined in Eq.~\eqref{e.coefs}. By the definition of $R$ in Sec.~\ref{s.effective}, the flux of $x$, identified with the physical flux in Eq.~\eqref{e.fluxid}, is
\begin{equation}
\vec J_x = R\,\vec J_{\tilde x} = -R\, D_{\tilde x}\nabla \tilde x.
\label{e.JxR}
\end{equation}
Let $y$ be the driving quantity of each process ($T$ for heat conduction, $v_\perp$ for viscosity, and $\rho_D$ for self-diffusion) and $\zeta$ the corresponding transport coefficient, so that $\vec J_x=-\zeta\nabla y$. A gradient of the driving quantity corresponds to a gradient of the ideal-form density, $\nabla\tilde x = K_y\nabla y$, where $K_y$ is a factor that, like $\tilde x$, is the same in the real and ideal systems; any departure is, by definition, included in $R$. Then $\zeta = R\,D_{\tilde x}K_y$ and, since $R=1$ in the ideal system, $\zeta_\text{id}=D_{\tilde x}^\text{id}\, K_y$. The factor $K_y$ cancels in the ratio:
\begin{equation}
\frac{\zeta}{\zeta_\text{id}} = \frac{D_{\tilde x}}{D^\text{id}_{\tilde x}}\, R,
\label{e.sigma}
\end{equation}
where $D_{\tilde x}/D^\text{id}_{\tilde x}$ is given by Eq.~\eqref{e.Dratio2} for heat conduction and by Eq.~\eqref{e.DratioU} for viscosity and self-diffusion. In terms of the specific heat along the path of constant $\tilde x$, Eq.~\eqref{e.ca}, the result for heat conduction follows from Eq.~\eqref{e.Dratio}:
\begin{equation}
\frac{\lambda}{\lambda_\text{id}} = \frac{2}{5}\left[c_{\tilde x_u} - T\left(\frac{\partial c_{\tilde x_u}}{\partial T}\right)_{\tilde x_u}\right] R_u,
\label{e.masterT}
\end{equation}
where $R_u$ is the departure factor for energy (Sec.~\ref{s.lambdaeta}). Equivalently,
\begin{equation}
\frac{\lambda}{\lambda_\text{id}} = \frac{\left(\partial^2 s/\partial\rho^2\right)_{\tilde x_u}}{\left(\partial^2 s_\text{id}/\partial\rho^2\right)_{\tilde x_u}}\, R_u.
\label{e.masterTs}
\end{equation}
This form follows from Appendix~\ref{a.hessian}. Since $-1/(\partial^2 s/\partial\rho^2)_{\tilde x}$ is proportional to the variance of the density of a cell at constant $\tilde x$, the ratio in Eq.~\eqref{e.masterTs} is the ratio between the density fluctuations at constant $\tilde x_u$ of the ideal and real systems.

For viscosity and self-diffusion, Eq.~\eqref{e.DratioU} gives
\begin{equation}
\frac{\zeta}{\zeta_\text{id}} = \frac{c_\rho}{c_{\tilde x}^\text{id}}\;\frac{a c_{\tilde x} - T\left(\frac{\partial c_{\tilde x}}{\partial T}\right)_{\tilde x}}{a c_\rho - T\cvT}\; R,
\label{e.master}
\end{equation}
or, equivalently,
\begin{equation}
\frac{\zeta}{\zeta_\text{id}} = \frac{\left(\partial^2 s/\partial\rho^2\right)_{\tilde x}}{\left(\partial^2 s_\text{id}/\partial\rho^2\right)_{\tilde x}}\;\frac{a c_\rho}{a c_\rho - T\cvT}\; R .
\label{e.masters}
\end{equation}
The dilute-gas value $\zeta_\text{id}$, obtained from the Chapman-Enskog solution of the Boltzmann equation \cite{chapman}, contains all the dynamical information.

Equations~\eqref{e.sigma}--\eqref{e.masters} follow from the theory of Sec.~\ref{s.ttt}, with the hypothesis of Eq.~\eqref{e.hyp}, and from the definition of $R$. Each transport coefficient is then determined by two inputs: the exponent $a$, which fixes the ratio $D_{\tilde x}/D^\text{id}_{\tilde x}$ through the Hessian, and the factor $R$, which is obtained in Secs.~\ref{s.fluct}--\ref{s.D} from hypotheses on the amounts exchanged.

\section{Equilibrium fluctuations of a fluid element}
\label{s.fluct}

For the three transport processes, the factor $R$ is obtained from equilibrium fluctuations. Instead of the cell of fixed volume of Sec.~\ref{s.cells}, it is convenient to consider here a fluid element that contains a fixed number of particles $N$ and exchanges energy and volume with its surroundings, because its fluctuations separate into a thermal and a mechanical part; the word cell is kept for this element. The relative volume fluctuations of the element are equal to the relative fluctuations of the number of particles of a cell of fixed volume, $\langle\delta V^2\rangle/V^2=\langle\delta N^2\rangle/N^2=1/(N\Gamma)$, where $\Gamma$ is the thermodynamic factor. The ensemble of the fluid element is the isothermal-isobaric one, in which the relevant fluctuations are those of the volume $V$ and of the enthalpy $\mathcal H$ of the cell \cite{callen,reichl}:
\begin{equation}
\langle\delta V^2\rangle = TV\kappa_T,\ 
\langle\delta V\,\delta \mathcal H\rangle = T^2V\alpha_P,\ 
\langle\delta \mathcal H^2\rangle = T^2N c_P,
\label{e.fluctNPT}
\end{equation}
where $\kappa_T$ is the isothermal compressibility, $\alpha_P$ the thermal expansion coefficient, and $c_P$ the specific heat per particle at constant pressure. The enthalpy fluctuations can be decomposed into a part correlated with the volume fluctuations, which is their projection onto $\delta V$, and a part uncorrelated with them, whose variance is the conditional variance at fixed volume:
\begin{align}
\frac{\langle\delta V\,\delta \mathcal H\rangle^2}{\langle\delta V^2\rangle} &= T^2N\,(c_P-c_\rho), \label{e.mech}\\
\langle\delta \mathcal H^2\rangle - \frac{\langle\delta V\,\delta \mathcal H\rangle^2}{\langle\delta V^2\rangle} &= T^2N\,c_\rho, \label{e.therm}
\end{align}
where we used $c_P-c_\rho = T\alpha_P^2/(\rho\kappa_T)$. These are called the mechanical and thermal parts of the enthalpy fluctuations, respectively. In the monatomic ideal gas, $c_\rho^\text{id}=3/2$ and $c_P^\text{id}-c_\rho^\text{id}=1$.

The mechanical part of an enthalpy fluctuation is
\begin{equation}
\delta\mathcal H_\text{mec} = \frac{\langle\delta V\,\delta \mathcal H\rangle}{\langle\delta V^2\rangle}\,\delta V = P_\text{th}\,\delta V,
\qquad P_\text{th}= T\left(\frac{\partial P}{\partial T}\right)_\rho,
\label{e.Hmec}
\end{equation}
i.e., the product of the thermal pressure $P_\text{th}$ and the displacement of the cell boundary, $\delta V$. In terms of the EOS, $P_\text{th}=\rho T Z_\text{th}$, with
\begin{equation}
Z_\text{th} = \frac{\partial (ZT)}{\partial T},
\label{e.Zth}
\end{equation}
which is equal to one in the ideal gas and to $Z$ for HS. The thermal pressure differs from the pressure by the internal pressure, $P_\text{th}-P=(\partial U/\partial V)_{T,N}$; for an EOS of the van der Waals type, the attractive term cancels in $P_\text{th}$, which is then the pressure of the repulsive reference fluid. The quantity $\Gamma_\text{th}$ introduced in Eq.~\eqref{e.Dratio2} is the thermodynamic factor associated with the thermal pressure,
\begin{equation}
\Gamma_\text{th} = \frac{1}{T}\left(\frac{\partial P_\text{th}}{\partial\rho}\right)_T = \frac{\partial (\rho Z_\text{th})}{\partial\rho} = \frac{\partial(\Gamma T)}{\partial T},
\label{e.Gth}
\end{equation}
which is equal to one in the ideal gas and to $\Gamma$ for HS. Unlike $\Gamma$, which vanishes at the critical point, $\Gamma_\text{th}$ remains of order one in the fluid region.

At fixed $N$, the energy of the cell changes as $\delta U=Nc_\rho\,\delta T+(P_\text{th}-P)\,\delta V$, so that $\delta\mathcal H=\delta U+P\,\delta V=Nc_\rho\,\delta T+P_\text{th}\,\delta V$. Since the fluctuations of the temperature and of the volume of the cell are statistically independent, $\langle\delta T\,\delta V\rangle=0$ \cite{landau}, the two terms are the thermal and mechanical parts: $\delta \mathcal{H}_\text{therm} = Nc_\rho\,\delta T$ and $\delta \mathcal{H}_\text{mec}=P_\text{th}\,\delta V$ as in  \eqref{e.Hmec}. The thermal part does not depend on the fluctuations of the volume, or density, of the cell, and the mechanical part does not depend on the fluctuations of its temperature; both coefficients, $Nc_\rho$ and $P_\text{th}$, are functions of the equilibrium state.

I assign to each transport process the part of the fluctuations associated with its mechanism of exchange. Energy is exchanged as heat, at fixed volume, and its amplitude is given by the thermal part, Eq.~\eqref{e.therm}. Momentum is transmitted through the forces acting across the cell boundaries, and its amplitude is described in terms of the thermal pressure that appears in the mechanical part, Eq.~\eqref{e.Hmec}. Composition is carried by particles, whose displacement across the boundaries is coupled to the volume fluctuations; I use the mechanical part, partitioned between the two labeled components. The volume fluctuations entering these expressions are the equilibrium fluctuations of a cell, which are present independently of the macroscopic perturbation: although a shear perturbation, for example, does not change the volume of the cells on average, the elementary exchanges take place across fluctuating cell boundaries.

\section{Thermal conductivity and viscosity}
\label{s.lambdaeta}

Energy and momentum are exchanged between particles in collisions. For these quantities, the factor $R$ represents the amplitude of the amount exchanged in the elementary events relative to its ideal value.

\subsection{Thermal conductivity}
\label{s.lambda}

Heat conduction is the transport of internal energy at uniform density. I identify the amplitude of the energy exchanged in the elementary events with the amplitude of the thermal part of the enthalpy fluctuations, Eq.~\eqref{e.therm}. Relative to the ideal gas, at the same density and temperature,
\begin{equation}
R_u = \sqrt{\frac{c_\rho}{c^\text{id}_\rho}}.
\label{e.Ru}
\end{equation}
The same factor is obtained from the energy fluctuations of a cell at fixed volume in the canonical ensemble, as in Ref.~\cite{alesLJ}. With Eq.~\eqref{e.Dratio2}, Eq.~\eqref{e.sigma} gives
\begin{equation}
\frac{\lambda}{\lambda_\text{id}} = \frac{2}{5}\left[c_\rho - T\cvT + 2\rho\cvr + \Gamma_\text{th}\right]\sqrt{\frac{c_\rho}{c^\text{id}_\rho}},
\label{e.lambda}
\end{equation}
which is the result of Ref.~\cite{alesLJ}, written there in terms of the specific heat at constant $\rho T$. Its predictions were tested against MD results for the LJ and the truncated and shifted LJ fluids in Ref.~\cite{alesLJ}; in this work, the comparison with simulation data is restricted to viscosity and self-diffusion. For HS, Eq.~\eqref{e.lambda} reduces to $\lambda/\lambda_\text{id}=(2\Gamma+3)/5$ \cite{hoyuelosHS}.

\subsection{Viscosity}
\label{s.eta}

The amplitude of the momentum exchanged between neighboring cells cannot be obtained from the momentum fluctuations of the particles: by equipartition, the variance of the momentum per particle is $mT$ in any classical fluid, and this contribution is already contained in $\tilde x_\perp$. What the interactions add is the collisional transfer of momentum, i.e., the momentum transmitted through interparticle forces, which dominates the momentum flux in a dense fluid.

A pressure is a momentum flux, i.e., momentum per unit area and unit time. I take the thermal pressure of Eq.~\eqref{e.Hmec} as the scale of the momentum flux across the boundary of a cell: it reduces to the kinetic flux $\rho T$ in the ideal gas and, as discussed in Sec.~\ref{s.fluct}, is determined mainly by the repulsive part of the interactions. Neither $P_\text{th}$ nor the volume fluctuation $\delta V$ selects a direction: the thermal pressure is the momentum flux across a surface of any orientation, and $\delta V$ corresponds to the displacement of the whole boundary of the cell. They fix a scale of the momentum exchanged that, in an isotropic fluid in equilibrium, is the same for all its components, including the one transverse to the wave vector, which is transported in viscosity; this component vanishes on average, but its amplitude does not. The momentum transferred across a portion of the boundary of area $A$ in an elementary exchange is the momentum flux times $A$ times the duration of the exchange. I estimate this duration as the time in which the boundary is displaced by $\delta\ell=\delta V/A$ at the thermal velocity $v_T=\sqrt{T/m}$, which, by the Maxwell distribution, is the same in the real and ideal systems. The momentum exchanged is then of the order of
\begin{equation}
P_\text{th}\,A\,\frac{\delta\ell}{v_T} = \frac{P_\text{th}\,\delta V}{v_T},
\label{e.pexch}
\end{equation}
i.e., the mechanical part of the enthalpy fluctuations, Eq.~\eqref{e.Hmec}, divided by the thermal velocity. Since $v_T$ is the same in both systems, the ratio of the momenta exchanged in the real and ideal systems is the ratio of the mechanical energies $P_\text{th}\,\delta V$. For the amplitude of the displacement of the boundary, I assume that, in a collisional exchange, it is resisted by the thermal pressure, so that it is that of a fluid with the EOS $P_\text{th}(\rho,T)$: $\langle\delta V^2\rangle_\text{th}/\langle\delta V^2\rangle_\text{id}=1/\Gamma_\text{th}$, with $\Gamma_\text{th}$ given by Eq.~\eqref{e.Gth}. Then,
\begin{equation}
R_\perp = \frac{P_\text{th}}{\rho T}\,\sqrt{\frac{\langle\delta V^2\rangle_\text{th}}{\langle\delta V^2\rangle_\text{id}}} = \frac{Z_\text{th}}{\sqrt{\Gamma_\text{th}}}.
\label{e.Rperp}
\end{equation}
For HS, $Z_\text{th}=Z$ and $\Gamma_\text{th}=\Gamma$, and Eq.~\eqref{e.Rperp} reduces to $R_\perp=Z/\sqrt\Gamma$, the factor obtained in Ref.~\cite{hoyuelosHS} from a mechanical-work hypothesis involving the pressure. For an EOS of the van der Waals type, $R_\perp$ is the same factor evaluated for the repulsive reference fluid. A factor built from $P$ would become small in liquid states where the pressure is small, and one built from $\Gamma$ would diverge at the critical point; $R_\perp$ remains positive and finite in every stable state.

With $a=1/2$ and Eq.~\eqref{e.DratioU}, Eq.~\eqref{e.sigma} gives
\begin{equation}
\frac{\eta}{\eta_\text{id}} = c_\rho\;\frac{c_\rho/2 - T\cvT + \rho\cvr + \Gamma_\text{th}/4}{c_\rho - 2T\cvT}\;\frac{Z_\text{th}}{\sqrt{\Gamma_\text{th}}}.
\label{e.eta}
\end{equation}
For HS, it reduces to $\eta/\eta_\text{id}=(\Gamma+3)Z/(4\sqrt\Gamma)$ \cite{hoyuelosHS}.

\section{Self-diffusion coefficient}
\label{s.D}

Composition differs from energy and momentum in that labels are not exchanged in collisions but carried by the particles, and the value of the label per particle is fixed. Neither the exponent $a_D$ nor the factor $R_D$ can therefore be associated with the thermal scale or with the fluctuation amplitude of the label itself. Both are obtained from a balance between the chemical-potential difference of the labeled components and the mechanical part of the enthalpy fluctuations: in the dilute limit, together with the first law for the transferred portion of fluid, the balance determines $a_D$; at finite density, it determines $R_D$.

\subsection{Mechanical balance for the transported composition}
\label{s.balance}

Since the particles are mechanically identical, I partition the mechanical part of the enthalpy fluctuations of a cell, Eq.~\eqref{e.Hmec}, between the labeled components in proportion to their numbers, $\delta\mathcal H_{\text{mec},A}=(N_A/N)\,P_\text{th}\,\delta V$ and $\delta\mathcal H_{\text{mec},B}=(N_B/N)\,P_\text{th}\,\delta V$. When a cell with a composition imbalance changes its density by $d\rho$ at a fixed number of particles, its volume changes by $\delta V=-N\,d\rho/\rho^2$, and the difference between the mechanical parts of the two components, per particle, is
\begin{equation}
\frac{\delta\mathcal H_{\text{mec},A}-\delta\mathcal H_{\text{mec},B}}{N} = -\frac{\rho_D P_\text{th}}{\rho^3}\,d\rho.
\label{e.HmecD}
\end{equation}
The two labeled components therefore receive different amounts of mechanical energy when the cell expands or contracts. We apply this result to the small portion of fluid that is transferred in a diffusion event. During the event, the state of the portion departs slightly from the equilibrium state of the fluid, at temperature $T_0$. We denote by $T$ the temperature of the portion, which, like that of a cell in Sec.~\ref{s.effective}, differs from $T_0$ during the event and tends to it at its end, and we use $\rho$ also for the density of the portion. For a given equilibrium state, the changes of the state of the portion define a local process $\mathcal P$ in the $(\rho,T)$ plane. In what follows, the differentials refer to changes along $\mathcal P$, and the thermodynamic coefficients, such as $P_\text{th}$, are evaluated at the equilibrium state, which is correct to first order in the departures. I assume that the difference in Eq.~\eqref{e.HmecD} determines the change of the chemical potential $\mu_D$, associated with $\rho_D$, of the portion. The composition density of the transferred portion is not the field $\rho_D$, whose equilibrium value vanishes, but the characteristic density $x_D$ of Sec.~\ref{s.effective}, whose dependence on the density and temperature is determined in Secs.~\ref{s.aD} and \ref{s.RD}. For small imbalances, Eq.~\eqref{e.muD} gives $\mu_D\simeq T\rho_D/\rho$, which for the transferred portion is $T x_D/\rho$. With the same replacement in Eq.~\eqref{e.HmecD}, the balance reads
\begin{equation}
d\left(\frac{T x_D}{\rho}\right) = -\frac{x_D P_\text{th}}{\rho^3}\,d\rho .
\label{e.balance}
\end{equation}
The balance is homogeneous in $x_D$: it determines the dependence of $x_D$ on the state of the portion but not its magnitude, which is contained in the constant $\epsilon_D$. For hard spheres, where $P_\text{th}=P$, it coincides with the mechanical-work hypothesis of Ref.~\cite{hoyuelosHS}, in which the work done on each labeled component is assigned in proportion to its number. Using $P_\text{th}$ instead of $P$ is what distinguishes the present formulation for fluids with attractive interactions, as shown in Sec.~\ref{s.RD}.

\subsection{Exponent}
\label{s.aD}

The exponent $a_D$ is obtained from the evolution of a small portion of the dilute gas that carries a composition imbalance in a diffusion event. During its lifetime, the portion follows the process $\mathcal P$ in the $(\rho,T)$ plane, which is determined by two conditions.

The first condition is the balance~\eqref{e.balance}. In the dilute gas, the composition density of the portion is $x_D=\tilde x_D$ and $P_\text{th}=\rho T_0$, and Eq.~\eqref{e.balance} reduces, to first order in the departures, to
\begin{equation}
d(T\tilde x_D) = 0
\label{e.pD}
\end{equation}
along $\mathcal P$. With $\tilde x_D\propto\rho\,T^{a_D}$, it gives
\begin{equation}
\left(\frac{d\ln\rho}{d\ln T}\right)_{\mathcal P} = -(1+a_D).
\label{e.slope1}
\end{equation}

The second condition is the first law for the portion. A portion with a composition imbalance on a length scale $k^{-1}$ remains identifiable during a time of the order of $\tau_D=(D_\text{id}k^2)^{-1}$, after which its labels are mixed with those of the surroundings. Since the portion exists as such only during this time, its departure from equilibrium, including that of its temperature, disappears on the same time scale, so that $dT/dt=-D_\text{id}k^2(T-T_0)$. This is not a property of the temperature field, which relaxes by conduction at its own rate, but of the state assigned to the portion during the event. Meanwhile, the portion exchanges heat with the surroundings by conduction on the same length scale, at the rate $\delta q/dt = -\lambda_\text{id}k^2(T-T_0)/\rho$ per particle. Eliminating time, the heat received per particle is $\delta q=\lambda_\text{id}\,dT/(\rho D_\text{id})$; only the ratio of the two rates enters, and the length scale $k^{-1}$ cancels. The first law per particle, including the work of compression, is $c^\text{id}_\rho\,dT=\delta q+(T_0/\rho)\,d\rho$, so that
\begin{equation}
\left(\frac{d\ln\rho}{d\ln T}\right)_{\mathcal P} = c^\text{id}_\rho-\frac{\lambda_\text{id}}{\rho D_\text{id}}.
\label{e.slope2}
\end{equation}

Comparing Eqs.~\eqref{e.slope1} and \eqref{e.slope2}, and using $c^\text{id}_P=c^\text{id}_\rho+1$, we obtain
\begin{equation}
a_D = \frac{\lambda_\text{id}}{\rho D_\text{id}} - c^\text{id}_P = c^\text{id}_P\,(\text{Le}_\text{id}-1),
\label{e.aD}
\end{equation}
where $\text{Le}_\text{id}=\lambda_\text{id}/(\rho c_P^\text{id} D_\text{id})$ is the Lewis number of the dilute gas, the ratio of the thermal diffusivity to the diffusion coefficient. The exponent is positive when heat diffuses faster than the labels. For $\text{Le}_\text{id}=1$, the heat received during the lifetime of the portion is that of an isobaric heating, $\rho\,T$ is constant along $\mathcal P$, and $a_D=0$.

In the first Chapman-Enskog approximation, the dilute-gas coefficients are given in terms of the reduced collision integrals $\Omega^{(1,1)}$ and $\Omega^{(2,2)}$, and $\lambda_\text{id}/(\rho D_\text{id}) = \frac{25}{8}\Omega^{(1,1)}/\Omega^{(2,2)}$ \cite{chapman,Hirschfelder}. Then $\text{Le}_\text{id}=\frac{5}{4}\Omega^{(1,1)}/\Omega^{(2,2)}$ for a monatomic gas ($c^\text{id}_P=5/2$) and
\begin{equation}
a_D = \frac{5}{8} + \frac{25}{8}\left(\frac{\Omega^{(1,1)}}{\Omega^{(2,2)}}-1\right).
\label{e.aDOmega}
\end{equation}
For HS, $\Omega^{(1,1)}=\Omega^{(2,2)}=1$ and $a_D=5/8$, the value obtained in Ref.~\cite{hoyuelosHS}. For the LJ potential, the ratio of collision integrals depends on temperature, and so does $a_D$. The exponent is a property of the equilibrium state: the dilute-gas coefficients in Eq.~\eqref{e.aD} are evaluated at $T_0$, while the process $\mathcal P$ describes the changes of the temperature $T$ of the portion, so that $\tilde x_D=\epsilon_D\,\rho\,T^{a_D(T_0)}$ along $\mathcal P$, as in Sec.~\ref{s.effective}. The Hessian involves the same kind of variation: its derivatives are taken with respect to the temperature of a cell that fluctuates around $T_0$, while $a_D(T_0)$ remains fixed.

\subsection{Departure factor}
\label{s.RD}

At finite density, we write $x_D=\tilde x_D R_D$. The factor $R_D$ is obtained from the balance~\eqref{e.balance}, i.e., from the mechanical part of the enthalpy fluctuations, which depends on the changes of the density of the portion but not on those of its temperature, with the coefficient $P_\text{th}$ evaluated at the equilibrium state (Sec.~\ref{s.fluct}). The departure of the temperature $T$ of the portion from $T_0$ enters only through the thermal part, which, for composition, is represented by the ideal-form density $\tilde x_D$, as in the dilute gas. Therefore, $R_D$ depends on the density of the portion, but not on its temperature: $R_D=R_D(\rho,T_0)$, where, as for $a_D$, $T_0$ is the temperature of the equilibrium state. With $P_\text{th}=\rho T_0 Z_\text{th}$, Eq.~\eqref{e.balance} gives
\begin{equation}
d\ln\left(\frac{T\tilde x_D}{\rho}\right) + d\ln R_D = -Z_\text{th}\,\frac{d\rho}{\rho}.
\end{equation}
Along $\mathcal P$, the ideal part satisfies Eq.~\eqref{e.pD}, i.e., $d\ln(T\tilde x_D/\rho)=-d\rho/\rho$, and $R_D$ changes only through the density, so that
\begin{equation}
\left(\frac{\partial\ln R_D}{\partial\rho}\right)_{T_0} = -\frac{Z_\text{th}-1}{\rho},
\label{e.dRD}
\end{equation}
with $Z_\text{th}$ evaluated at $(\rho,T_0)$. Applied to each equilibrium state at temperature $T_0$, Eq.~\eqref{e.dRD} determines the dependence of $R_D$ on density, and it is integrated from the dilute limit, where $R_D=1$. Since the temperature of the portion no longer appears, the subscript of $T_0$ is dropped from here on, and the final results are functions of the equilibrium state denoted by $(\rho,T)$. The Maxwell relation $(\partial s'/\partial\rho)_T=-\rho^{-2}(\partial P/\partial T)_\rho = -Z_\text{th}/\rho$ for the entropy per particle gives, for its excess part,
\begin{equation}
\left(\frac{\partial s'_\text{ex}}{\partial\rho}\right)_T = -\frac{Z_\text{th}-1}{\rho},
\end{equation}
with $s'_\text{ex}\to 0$ for $\rho\to 0$. Therefore, combining with Eq.\ \eqref{e.dRD}, we have
\begin{equation}
R_D = e^{s'_\text{ex}},
\label{e.RD}
\end{equation}
where $s'_\text{ex}$ is the excess entropy per particle. It is obtained from the EOS through
\begin{equation}
s'_\text{ex} = \frac{u'_\text{ex}-f'_\text{ex}}{T},\qquad
\frac{f'_\text{ex}}{T} = \int_0^\rho \frac{Z(\bar\rho,T)-1}{\bar\rho}\,d\bar\rho,
\label{e.sex}
\end{equation}
where $f'_\text{ex}$ is the excess Helmholtz free energy per particle and $u'_\text{ex}=-T^2\,\partial(f'_\text{ex}/T)/\partial T$ is the excess internal energy per particle.

The use of the thermal pressure is essential for this result. With $P$ instead of $P_\text{th}$ in Eq.~\eqref{e.balance}, the same procedure gives $\partial\ln R_D/\partial\rho=-(Z-1)/\rho$ and $R_D=e^{-f'_\text{ex}/T}=e^{s'_\text{ex}}e^{-u'_\text{ex}/T}$, the factor obtained in Ref.~\cite{hoyuelosHS}. Both expressions coincide for HS, where $u'_\text{ex}=0$, but differ for a fluid with attractive interactions: in a dense liquid, $u'_\text{ex}<0$ and the energetic factor $e^{-u'_\text{ex}/T}$ is much larger than one, so that the interactions would enhance self-diffusion with respect to the dilute gas, contrary to observation. The difference between the two factors comes from the energetic contribution to the pressure, $P-P_\text{th}=-(\partial U/\partial V)_{T,N}$, which does not enter the mechanical part of the enthalpy fluctuations. The factor $e^{s'_\text{ex}}$ is the ratio, per particle, of the numbers of accessible microstates in the real and ideal systems at the same density and temperature; it is smaller than one since $s'_\text{ex}<0$ for the states considered, and it connects the present result with the excess-entropy scaling of transport coefficients \cite{rosenfeld,rosenfeld3,dyre}.

Using Eq.~\eqref{e.DratioU} with $a=a_D$, Eq.~\eqref{e.sigma} gives
\begin{equation}
\frac{D}{D_\text{id}} = \frac{c_\rho}{c_\rho^\text{id}+a_D}\;\frac{a_D c_\rho - T\cvT + 2a_D\rho\cvr + a_D^2\,\Gamma_\text{th}}{a_D c_\rho - T\cvT}\;e^{s'_\text{ex}}.
\label{e.D}
\end{equation}
For HS, with $a_D=5/8$, it reduces to $D/D_\text{id}=(5\Gamma+12)\,e^{s'_\text{ex}}/17$ \cite{hoyuelosHS}. Equation~\eqref{e.D} requires $a_D>0$, i.e., $\text{Le}_\text{id}>1$: for $a_D\to 0$, $\tilde x_D$ becomes proportional to $\rho$ and can no longer be used together with $\rho$ as a thermodynamic coordinate. With the collision integrals of Refs.~\cite{fokin,meier1}, Eq.~\eqref{e.aDOmega} gives $0.31\le a_D\le 0.36$ for $0.7\le T^*\le 10$ for the LJ fluid.

\section{Comparison with numerical results}
\label{s.comp}

I compare the predictions of Eqs.~\eqref{e.eta} and \eqref{e.D} with the MD results of Meier et al.\ for the shear viscosity and the self-diffusion coefficient of the LJ fluid \cite{meier1,meier3,meier2}, obtained with a cutoff radius $r_c$ between 5 and 6.5 in LJ units, which is large enough to represent the full LJ potential. We use LJ reduced units, $\rho^*=\rho\sigma_\text{LJ}^3$, $T^*=T/\varepsilon$, $\eta^*=\eta\,\sigma_\text{LJ}^2/\sqrt{m\varepsilon}$, and $D^*=D\sqrt{m/\varepsilon}/\sigma_\text{LJ}$, where $\sigma_\text{LJ}$ and $\varepsilon$ are the parameters of the LJ potential and $m$ is the mass of a particle. The dilute-gas values are given by the first Chapman-Enskog approximation \cite{chapman,Hirschfelder},
\begin{equation}
\eta^*_\text{id}=\frac{5}{16}\frac{\sqrt{T^*/\pi}}{\Omega^{(2,2)}}, \qquad \rho^* D^*_\text{id}=\frac{3}{8}\frac{\sqrt{T^*/\pi}}{\Omega^{(1,1)}},
\label{e.dilute}
\end{equation}
with the reduced collision integrals of Refs.~\cite{fokin,meier1}, which are also used in Eq.~\eqref{e.aDOmega} for $a_D$.

I consider only supercritical temperatures, $1.5\le T^*\le 6$; the critical temperature of the LJ fluid is $T_c^*\simeq 1.32$ \cite{thol}. The theoretical results involve derivatives of the free energy up to third order and are therefore very sensitive to the changes of curvature associated with the critical point, whose effect is already visible at the lowest temperatures considered (see below). The theory applies to a homogeneous, thermodynamically stable fluid. In the figures for TTT, the region beyond the freezing line is shown in gray, as in Ref.~\cite{alesLJ}: for each temperature, the theoretical isotherm crosses the freezing line at the freezing density $\rho_\text{f}^*(T^*)$, obtained from the solid-liquid coexistence data of Pedersen \cite{pedersen2013}, and the gray region lies on the high-density side of the parametric curve formed by these crossing points. The thermodynamic quantities are obtained from the EOS of Thol et al.\ \cite{thol}. The EOS of Kolafa and Nezbeda \cite{kolafa} and the SAFT-VR Mie EOS of Lafitte et al.\ \cite{lafitte} give similar results (see below). All the calculations of this section, including the simulation data, the EOS, and the evaluation of RET, are contained in a SageMath notebook available in Ref.~\cite{sagelink}.

The agreement with the simulation data is quantified by the mean absolute percentage error (MAPE) of each isotherm,
\begin{equation}
\text{MAPE}=\frac{100\%}{n}\sum_{i=1}^{n}\left|\frac{\zeta_{\text{th},i}-\zeta_{\text{MD},i}}{\zeta_{\text{MD},i}}\right|,
\label{e.MAPE}
\end{equation}
where $\zeta$ is $\eta$ or $D$, and the sum runs over the $n$ simulation points of the isotherm in the density range indicated below.

I compare the results with those of the revised Enskog theory (RET) for Mie fluids of Jervell and Wilhelmsen \cite{jervell,jervell2024predicting}, which contains no parameters fitted to transport data. In RET, the effect of density enters through the radial distribution function at contact, which increases the collision frequency, and through the instantaneous transfer of momentum and energy across a distance of the order of the particle size at the moment of the collision. The contact value of the radial distribution function is obtained from the perturbation theory underlying the SAFT-VR Mie EOS \cite{lafitte}. For self-diffusion of identical particles, the collisional transfer does not contribute, and, in the first Enskog approximation, $D/D_\text{id}$ is the inverse of the contact value of the radial distribution function; this result does not depend on the transfer lengths, so it is the same in the original version of the theory \cite{jervell} and in the more recent one \cite{jervell2024predicting}. For the viscosity, I use the more recent version, RET-EWCA \cite{jervell2024predicting}, in which the collisional contributions are evaluated with a momentum transfer length, obtained from the distance of closest approach of colliding particles weighted by the momentum exchanged in the collision and averaged over the collision parameters (exchange-weighted closest approach, EWCA).

\subsection{Self-diffusion}
\label{s.compD}

\begin{figure*}
\includegraphics[width=0.42\textwidth]{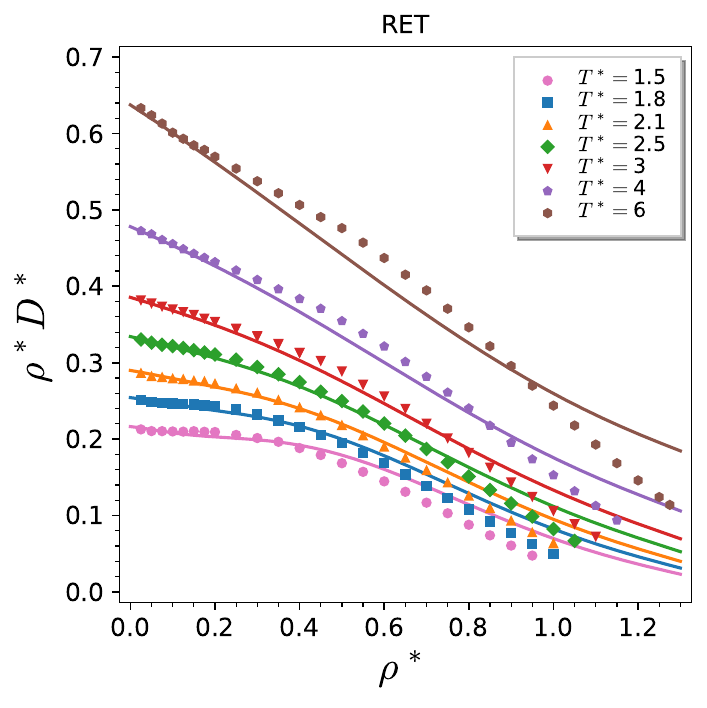}\hspace{0.04\textwidth}
\includegraphics[width=0.42\textwidth]{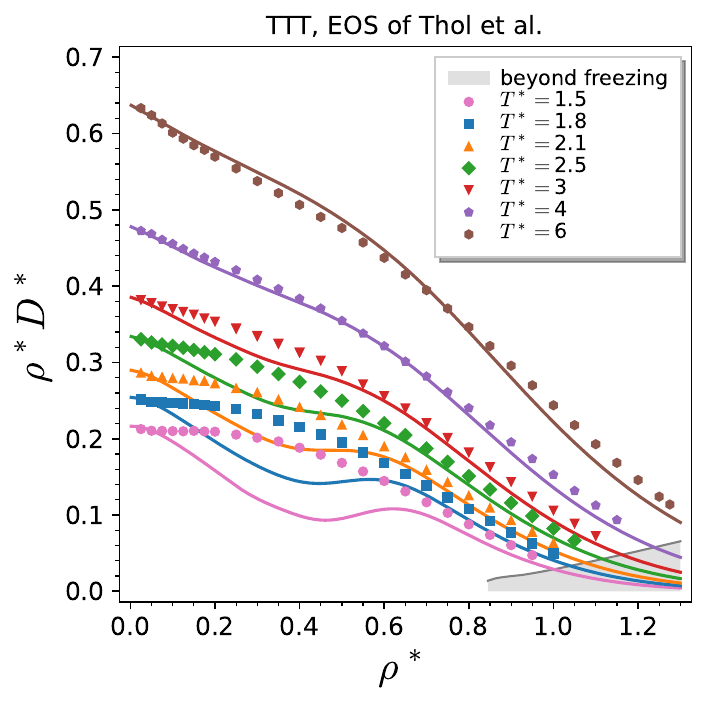}
\caption{$\rho^*D^*$ of the LJ fluid as a function of density for the supercritical temperatures indicated in the legends. Symbols: MD results of Meier et al.\ \cite{meier1,meier2}. (a, left panel) Lines: RET \cite{jervell}, with the contact value of the radial distribution function given by the SAFT-VR Mie EOS \cite{lafitte}. (b, right panel) Lines: present theory (TTT), Eq.~\eqref{e.D}, with the EOS of Thol et al.\ \cite{thol}; the gray region lies beyond the freezing line. In both panels, $D_\text{id}$ is given by Eq.~\eqref{e.dilute}.}
\label{f.D}
\end{figure*}

\begin{table}
\caption{MAPE (in \%), Eq.~\eqref{e.MAPE}, of RET and TTT (EOS of Thol et al.) for the self-diffusion coefficient with respect to the simulation results of Meier et al.\ \cite{meier1,meier2}, for $\rho^*\le 1.2$ (see Fig.~\ref{f.D}).}
\label{t.D}
\begin{ruledtabular}
\begin{tabular}{ccc}
$T^*$ & RET & TTT \\
\hline
1.5 & 13 & 23 \\
1.8 & 11 & 16 \\
2.1 & 7.5 & 12 \\
2.5 & 7.0 & 8.4 \\
3 & 6.6 & 6.1 \\
4 & 7.1 & 3.5 \\
6 & 7.0 & 3.7 \\
\end{tabular}
\end{ruledtabular}
\end{table}

The comparison for self-diffusion is shown in Fig.~\ref{f.D}, and the MAPE, for all the simulation data with $\rho^*\le1.2$, is given in Table~\ref{t.D}. All these points lie below the freezing density. Both theories tend to the dilute-gas value at low density, in agreement with the simulation data. RET predicts a smooth decrease of $\rho D$ with density; at high temperatures it underestimates the simulation data at intermediate densities and overestimates them in the dense fluid, with a MAPE of about 7\% for $T^*\ge2.1$, and between 11\% and 13\% at the two lowest temperatures. For $T^*\ge3$, TTT reproduces the shape of the isotherms, including the change of curvature at intermediate densities and the steep decrease in the dense fluid, with MAPE values between 3.5\% and 6.1\%, smaller than those of RET. As the temperature approaches the critical one, the TTT isotherms develop a minimum at intermediate densities, around $\rho^*\simeq0.4$, which is not present in the simulation data, and the MAPE increases up to 23\% at $T^*=1.5$. This is the effect of the critical point mentioned above: the thermodynamic factors in Eq.~\eqref{e.D}, in particular the derivatives of the specific heat, vary rapidly in this region. At high densities, the agreement of TTT with the simulation data is good at all temperatures. With the EOS of Kolafa and Nezbeda or the SAFT-VR Mie EOS, the errors of TTT are very similar: the MAPE differs by at most two percentage points from the values of Table~\ref{t.D}.

\subsection{Viscosity}
\label{s.competa}

\begin{figure*}
\includegraphics[width=0.42\textwidth]{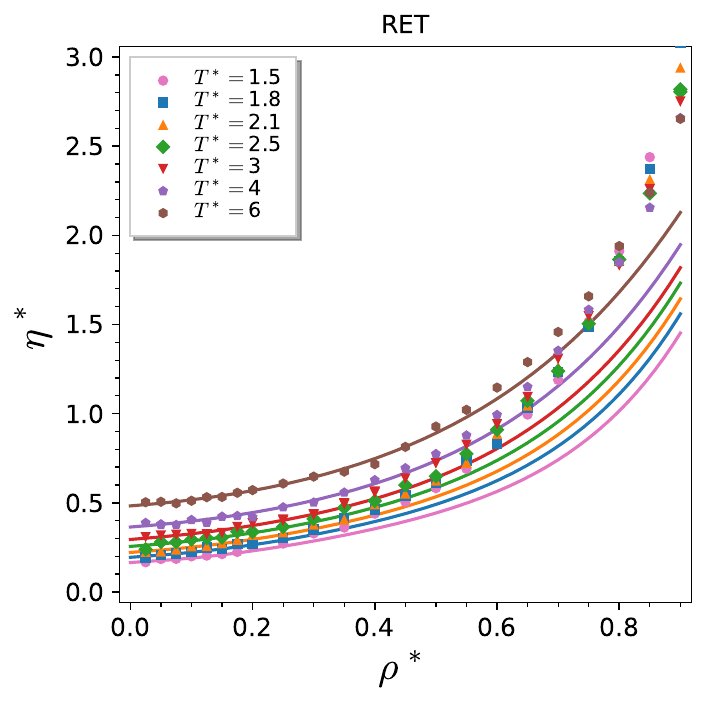}\hspace{0.04\textwidth}
\includegraphics[width=0.42\textwidth]{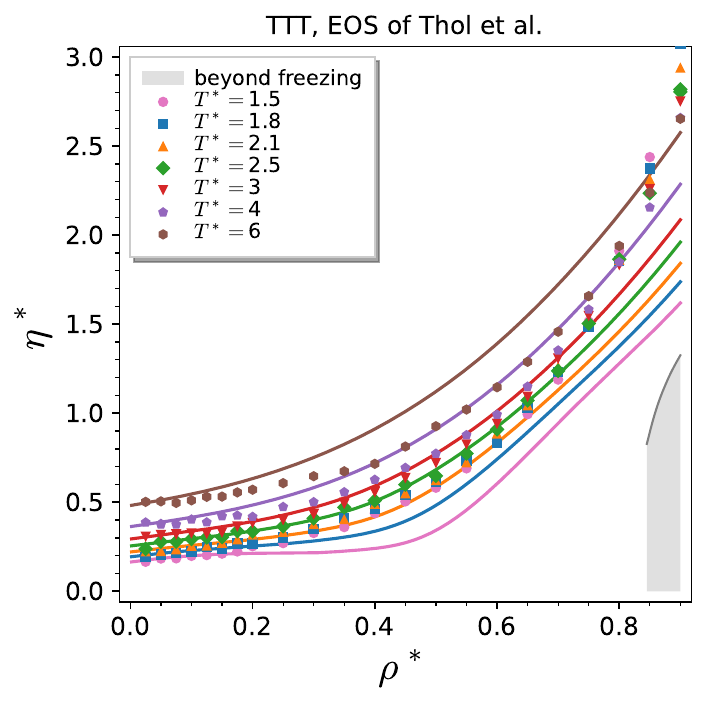}
\caption{Reduced shear viscosity $\eta^*$ of the LJ fluid as a function of density for the supercritical temperatures indicated in the legends. Symbols: MD results of Meier et al.\ \cite{meier1,meier3}. (a, left panel) Lines: RET-EWCA \cite{jervell2024predicting}, with the contact value of the radial distribution function given by the SAFT-VR Mie EOS \cite{lafitte}. (b, right panel) Lines: TTT, Eq.~\eqref{e.eta}, with the EOS of Thol et al.\ \cite{thol}; the gray region lies beyond the freezing line. In both panels, $\eta_\text{id}$ is given by Eq.~\eqref{e.dilute}.}
\label{f.eta}
\end{figure*}

\begin{table}
\caption{MAPE (in \%), Eq.~\eqref{e.MAPE}, of RET-EWCA and TTT for the shear viscosity with respect to the simulation results of Meier et al.\ \cite{meier1,meier3}, for $\rho^*\le0.8$ (see Fig.~\ref{f.eta}). TTT is evaluated with the EOS of Thol et al.\ \cite{thol}.}
\label{t.eta}
\begin{ruledtabular}
\begin{tabular}{ccc}
$T^*$ & RET & TTT \\
\hline
1.5 & 18 & 21 \\
1.8 & 15 & 13 \\
2.1 & 12 & 7.7 \\
2.5 & 11 & 3.6 \\
3 & 7.4 & 5.5 \\
4 & 5.5 & 10 \\
6 & 3.5 & 14 \\
\end{tabular}
\end{ruledtabular}
\end{table}

The comparison for the viscosity is shown in Fig.~\ref{f.eta} and Table~\ref{t.eta}. In this case, the MAPE is restricted to densities $\rho^*\le0.8$, because of the proximity of the freezing line: the viscosity increases steeply in the dense fluid, and, as for self-diffusion, the theory is sensitive to the changes of curvature of the free energy that precede the phase transition. RET-EWCA gives a good description at high temperatures, with a MAPE of 3.5\% at $T^*=6$, but it underestimates the viscosity at intermediate and high densities as the temperature decreases, and its MAPE increases up to 18\% at $T^*=1.5$, in agreement with the trend reported in Ref.~\cite{jervell2024predicting}. TTT with the EOS of Thol et al.\ gives its best results at intermediate temperatures, with a MAPE between 3.6\% and 7.7\% for $2.1\le T^*\le3$, smaller than that of RET-EWCA. At high temperatures, TTT overestimates the viscosity at intermediate densities, with a MAPE of 14\% at $T^*=6$, while at the lowest temperature, $T^*=1.5$, it underestimates it, as a consequence of the proximity of the critical point. With the EOS of Kolafa and Nezbeda or the SAFT-VR Mie EOS, the MAPE of TTT differs by about one percentage point or less from that obtained with the EOS of Thol et al., except at $T^*=1.5$, where the SAFT-VR Mie EOS gives 17\%.

As in the case of the thermal conductivity \cite{alesLJ}, the result depends on the EOS through the derivatives of the free energy. The differences between the results obtained with different EOS reflect the difficulty of describing simultaneously the thermodynamic properties and their higher-order derivatives. In a recent assessment of 20 EOS for the LJ fluid, Stephan et al.\ \cite{stephan2020} found that none of them meets two criteria at the same time: the absence of unphysical artifacts when used for extrapolations, and the description of the simulation data within their statistical uncertainty in most fluid regions. Empirical EOS, such as that of Thol et al., give the most accurate description of the thermal and caloric properties, including the heat capacities, but show artificial oscillations of the isotherms in the two-phase region, which extend up to the critical temperature in the case of the EOS of Thol et al. EOS based on perturbation theory, such as the SAFT-VR Mie EOS, are essentially free of these artifacts but are less accurate in the homogeneous fluid \cite{stephan2020}. Since TTT involves derivatives of the free energy up to third order, it is sensitive to both kinds of deficiency, especially close to the critical point.

\section{Summary of results}
\label{s.summary}

The theory gives each transport coefficient as the product of its dilute-gas value and a function of equilibrium thermodynamic properties. For the thermal conductivity, the shear viscosity, and the self-diffusion coefficient, Eqs.~\eqref{e.lambda}, \eqref{e.eta}, and \eqref{e.D} read
\begin{align*}
\frac{\lambda}{\lambda_\text{id}} &= \frac{2}{5}\left[c_\rho - T\cvT + 2\rho\cvr + \Gamma_\text{th}\right]\sqrt{\frac{c_\rho}{c^\text{id}_\rho}}, \\
\frac{\eta}{\eta_\text{id}} &= c_\rho\;\frac{c_\rho/2 - T\cvT + \rho\cvr + \Gamma_\text{th}/4}{c_\rho - 2T\cvT}\;\frac{Z_\text{th}}{\sqrt{\Gamma_\text{th}}}, \\
\frac{D}{D_\text{id}} &= \frac{c_\rho}{c_\rho^\text{id}+a_D}\;\frac{a_D c_\rho - T\cvT + 2a_D\rho\cvr + a_D^2\,\Gamma_\text{th}}{a_D c_\rho - T\cvT}\;e^{s'_\text{ex}}, 
\end{align*}
with $Z_\text{th}=\partial(ZT)/\partial T$, $\Gamma_\text{th}=\partial(\Gamma T)/\partial T$, and $a_D$ given by Eq.~\eqref{e.aDOmega}. The thermodynamic input consists of the specific heat at constant density and its first derivatives, $Z_\text{th}$, $\Gamma_\text{th}$, and the excess entropy, all of which are obtained from the Helmholtz free energy and its derivatives up to third order; the dynamical input consists of the dilute-gas values and, for self-diffusion, the ratio of collision integrals that determines $a_D$.

These results rest on the following hypotheses.

(i) \textit{Equal modification of the function $f$}, Eq.~\eqref{e.hyp}. The interactions modify the function $f$ of the exchanges of the transported quantity in the same way as that of the crossings of particles through the boundary of a cell, whose modification is known exactly, Eq.~\eqref{e.fN}. It represents a change of the time scale of the exchanges between two cells that does not depend on the quantity exchanged. 
Its justification is that it leads to a simple structure represented by Eq.~\eqref{e.main2}, in which the Onsager coefficient is given by the ratio of the determinants of the Hessians of the Massieu function $\Phi$, the potential of the ensemble in which $N$ is exchanged with a reservoir. In thermodynamic geometry, this Hessian corresponds to the representation obtained from the entropy by a partial Legendre transformation \cite{salamon,bravetti}; see Sec.~\ref{s.nontransp}.

(ii) \textit{Departure factors from equilibrium fluctuations}. The factor $R$ is interpreted as the ratio between the amplitudes of the amounts exchanged in the real and ideal systems, and it is obtained from the decomposition of the enthalpy fluctuations of a fluid element into a thermal and a mechanical part (Sec.~\ref{s.fluct}). Each process is assigned the part associated with its mechanism of exchange: the thermal part for energy, which is exchanged as heat at fixed volume, and the mechanical part for momentum, which is transmitted by the forces across the boundaries, and for composition, which is carried by the particles whose displacement is coupled to the volume fluctuations. The mechanical part is $P_\text{th}\,\delta V$, so that the thermal pressure, and not the pressure, sets the scale of the momentum and of the composition exchanged; it is the entropic part of the pressure, determined mainly by the repulsive interactions. For viscosity, the momentum exchanged is estimated as $P_\text{th}\,\delta V/v_T$, Eq.~\eqref{e.pexch}, and it is assumed that the displacement of the boundary in a collisional exchange is that of a fluid with the EOS $P_\text{th}(\rho,T)$, which gives Eq.~\eqref{e.Rperp}. For composition, the mechanical part is partitioned between the labeled components in proportion to their numbers, and the difference determines the change of the chemical potential associated with the composition of the portion of fluid transferred in a diffusion event, Eq.~\eqref{e.balance}; this balance gives $R_D=e^{s'_\text{ex}}$.

(iii) \textit{Ideal-form density for composition and dilute-gas process of a diffusion event}, Eq.~\eqref{e.xtilde} and Sec.~\ref{s.aD}. The form $\tilde x_D\propto\rho T^{a_D}$ is a hypothesis, with $a_D$ a property of the equilibrium state, evaluated at its temperature $T_0$ and therefore constant in the Hessian, where the temperature $T$ of the cell fluctuates. The exponent is determined by the process of a diffusion event in the dilute gas: the mechanical balance of (ii) reduces to $d(T\tilde x_D)=0$, where $T$ is the temperature of the transferred portion, and the portion of fluid relaxes to equilibrium on the time scale of diffusion, while it exchanges heat by conduction. These conditions give $a_D=c_P^\text{id}(\text{Le}_\text{id}-1)$, which reduces to the value $a_D=5/8$ obtained for HS in Ref.~\cite{hoyuelosHS} and is positive for the LJ potential.

The comparison with simulation results in Sec.~\ref{s.comp} tests these hypotheses jointly, together with the accuracy of the EOS used to evaluate the thermodynamic quantities.

For HS, $c_\rho=c_\rho^\text{id}$, $Z_\text{th}=Z$, $\Gamma_\text{th}=\Gamma$, and $a_D=5/8$, and the three expressions reduce to
\begin{align*}
\frac{\lambda}{\lambda_\text{id}} &= \frac{2\Gamma+3}{5}, \\
\frac{\eta}{\eta_\text{id}} &= \frac{(\Gamma+3)\,Z}{4\sqrt\Gamma}, \\
\frac{D}{D_\text{id}} &= \frac{5\Gamma+12}{17}\,e^{s'_\text{ex}}.
\end{align*}
These expressions were obtained and verified in Ref.~\cite{hoyuelosHS}.

\section{Conclusions}
\label{s.conclusions}

A unified formulation of the thermodynamic transport theory for the thermal conductivity, the shear viscosity, and the self-diffusion coefficient of simple fluids is presented. Each transport coefficient is the product of its dilute-gas value and a function of equilibrium thermodynamic properties, with no adjustable parameters. The three processes differ in the exponent $a$ and in the departure factor $R$ that appear in the effective transported density $x=\epsilon\, \rho T^a R$. The use of the thermal pressure, instead of the pressure, in the mechanical part of the enthalpy fluctuations is essential for fluids with attractive interactions, and it gives a factor that connects the result for self-diffusion with excess-entropy scaling. For the thermal conductivity, the result of Ref.~\cite{alesLJ} is recovered, and, for HS, the results of Ref.~\cite{hoyuelosHS}. 

The bulk viscosity is outside the present scheme. In TTT, a transport coefficient describes the diffusive relaxation of a conserved density between neighboring cells, and it is given by its dilute-gas value multiplied by a thermodynamic factor. The bulk viscous stress is also a momentum flux, but the longitudinal momentum is coupled to the density and propagates as sound instead of diffusing. Moreover, the bulk viscosity of a dilute monatomic gas vanishes, so that there is no ideal value to which a thermodynamic factor could be applied. A thermodynamic description of the bulk viscosity would require a different formulation.

The predictions for the viscosity and the self-diffusion coefficient were compared with MD results for the LJ fluid at supercritical temperatures, $1.5\le T^*\le6$, and with RET for Mie fluids, which, to my knowledge, is the only predictive kinetic theory of the transport coefficients of dense fluids. The recent REST framework \cite{jervell2026}, which extrapolates RET results to the liquid-like region through a residual-entropy scaling relation, also contains no parameters fitted to transport data, but it involves additional choices, such as the boundary of the region where RET is considered reliable and the functional form of the extrapolation; in TTT, the transport coefficients follow directly from the thermodynamic state, without extrapolation schemes \cite{alesLJ}. With the EOS of Thol et al., the MAPE of TTT for self-diffusion is between 3.5\% and 6.1\% for $T^*\ge3$, smaller than that of RET, and increases up to 23\% at $T^*=1.5$. For the viscosity, with densities $\rho^*\le0.8$, the MAPE of TTT is between 3.6\% and 7.7\% for $2.1\le T^*\le 3$, smaller than that of RET-EWCA, while RET-EWCA is more accurate at the highest temperatures. Overall, the accuracy of TTT is similar to that of RET, although the two approaches follow very different routes: in RET, the effect of density enters through the structure of the fluid, i.e., the radial distribution function at contact, and through the dynamics of binary collisions; in TTT, the dynamics enters only through the dilute-gas values, and the effect of density is given by thermodynamic quantities obtained from the free energy.

The largest errors of TTT appear at the temperatures closest to the critical point, where the derivatives of the specific heat vary rapidly and the result is very sensitive to the curvature of the free energy. For this reason, the comparison was restricted to supercritical temperatures and, for the viscosity, to densities below $\rho^*=0.8$, sufficiently far from the freezing line. The theory is sensitive to the choice of the EOS, especially near the critical point. As shown by the assessment of Stephan et al.\ \cite{stephan2020}, none of the available EOS for the LJ fluid combines the accuracy of the empirical EOS with the absence of artifacts of those based on perturbation theory. Since TTT only requires an EOS, I expect that an improvement in the description of the free energy and its derivatives, especially near the critical point, would translate into more accurate predictions of the transport coefficients.

\begin{acknowledgments}
This work was partially supported by the Universidad Nacional de Mar del Plata (UNMdP), Argentina, under Grant No. 80020250500008MP. The AI tool Claude Opus 5.5 (Anthropic) was used to draft and revise parts of the text and of the SageMath notebook \cite{sagelink}; the author reviewed the results and takes full responsibility for the content.
\end{acknowledgments}

\appendix

\section{Hessian matrix}
\label{a.hessian}

The calculation of the Hessian of $\phi$, Eq.~\eqref{e.phi}, with respect to its natural variables, is presented here, as required by Eq.~\eqref{e.main2}. We write its differential as
\begin{equation}
d\phi = \tilde Y\, d\tilde x + \rho\, d\mu'_T,\qquad \mu'_T = \mu_T + \frac{c_\rho}{a},
\end{equation}
with $\tilde Y$ given by Eq.~\eqref{e.Ytilde}. The Hessian with respect to the natural variables is $H_\phi=\partial(\tilde Y,\rho)/\partial(\tilde x,\mu'_T)$. Since all quantities are explicit functions of $(T,\rho)$, it is simpler to calculate the inverse, $H_\phi^{-1}=\partial(\tilde x,\mu'_T)/\partial(\tilde Y,\rho)$, using the chain rule. We need the thermodynamic relations
\begin{equation}
\frac{\partial \mu_T}{\partial T} = -\frac{1}{T}\cvrr, \qquad
\frac{\partial \mu_T}{\partial \rho} = \frac{1}{\rho}\Gamma_\text{th};
\label{e.relmu}
\end{equation}
the first follows from the equality of the mixed second derivatives of $s$ in Eq.~\eqref{e.dsTrho}, and the second from $\partial\mu/\partial\rho = T\Gamma/\rho$ and $\Gamma_\text{th} = \partial(T\Gamma)/\partial T$, where $\Gamma=\partial(\rho Z)/\partial\rho$ is the thermodynamic factor and $Z=P/(\rho T)$ is the compressibility factor. We also define
\begin{equation}
A_a = a c_\rho - T\cvT,
\end{equation}
so that $\partial\tilde Y/\partial T = -A_a/(\epsilon a T^{a+1})$. The result is
\begin{equation}
H_\phi^{-1} = \frac{1}{A_a}
\begin{pmatrix}
-\epsilon^2 a^2 \rho T^{2a} & \epsilon T^a\left(A_a + a\rho\cvr\right) \\[4pt]
\epsilon T^a\left(A_a + a\rho\cvr\right) & \frac{A_a}{\rho}\Gamma_\text{th} - \rho\left(\cvr\right)^2
\end{pmatrix},
\end{equation}
and its determinant is
\begin{equation}
\det H_\phi^{-1} = -\frac{\epsilon^2 T^{2a}}{A_a}\, B_a,
\qquad
B_a = A_a + 2a\rho\cvr + a^2\Gamma_\text{th}.
\label{e.detH}
\end{equation}
Since $ds=\tilde Y\,d\tilde x-\mu'_T\,d\rho$, the determinant can be written in terms of the second derivatives of $s$ with respect to $\tilde x$ and $\rho$. Using the chain rule for Jacobians, with $(\tilde x,\rho)$ as intermediate variables, the determinant factorizes as
\begin{equation}
\det H_\phi = \left(\frac{\partial\tilde Y}{\partial\tilde x}\right)_\rho\left(\frac{\partial\rho}{\partial\mu'_T}\right)_{\tilde x}
= -\frac{\left(\partial^2 s/\partial\tilde x^2\right)_\rho}{\left(\partial^2 s/\partial\rho^2\right)_{\tilde x}},
\label{e.detHs}
\end{equation}
i.e., minus the ratio of the curvatures of the entropy along $\tilde x$ at constant density and along the density at constant $\tilde x$, as in Eq.~\eqref{e.detHPhi}. The two second derivatives are
\begin{align}
\left(\frac{\partial^2 s}{\partial\tilde x^2}\right)_\rho &= \left(\frac{\partial\tilde Y}{\partial\tilde x}\right)_\rho = -\frac{A_a}{\epsilon^2a^2\rho T^{2a}},
\label{e.s2x}\\
\left(\frac{\partial^2 s}{\partial\rho^2}\right)_{\tilde x} &= -\left(\frac{\partial\mu'_T}{\partial\rho}\right)_{\tilde x} = -\frac{B_a}{a^2\rho},
\label{e.s2rho}
\end{align}
and their ratio reproduces Eq.~\eqref{e.detH}. In Eq.~\eqref{e.DratioL}, the first one cancels with $(\partial\tilde Y/\partial\tilde x)_\rho$, and only the curvature of the entropy along the density at constant $\tilde x$ remains. In the ideal case, $A_a^\text{id}=a c^\text{id}_\rho$ and $B^\text{id}_a=a(c^\text{id}_\rho+a)$. Therefore,
\begin{equation}
\frac{\det H^\text{id}_\phi}{\det H_\phi} = \frac{c^\text{id}_\rho}{c^\text{id}_\rho + a}\,\frac{B_a}{A_a},
\end{equation}
which, through Eq.~\eqref{e.main2}, gives Eq.~\eqref{e.Lx} once $B_a$ is expressed in terms of $c_{\tilde x}$, as shown next.

The specific heat along the path of constant $\tilde x$ is
\begin{equation}
c_{\tilde x} = T\left(\frac{\partial s'}{\partial T}\right)_\rho - a\rho\left(\frac{\partial s'}{\partial\rho}\right)_T = c_\rho + a\,\frac{\partial(ZT)}{\partial T},
\end{equation}
where we used that $d\rho/\rho=-a\,dT/T$ at constant $\tilde x=\epsilon\rho T^a$, and the Maxwell relation $(\partial s'/\partial\rho)_T=-\rho^{-2}(\partial P/\partial T)_\rho = -\rho^{-1}\partial(ZT)/\partial T$. The derivative along the same path is obtained with the operator
\begin{equation}
T\left(\frac{\partial}{\partial T}\right)_{\tilde x} = T\frac{\partial}{\partial T} - a\rho\frac{\partial}{\partial\rho}.
\end{equation}
Applying it to $c_{\tilde x}$ and using the identities
\begin{equation}
\rho\cvr = -T\,\frac{\partial^2 (ZT)}{\partial T^2},
\qquad
\Gamma_\text{th} = \frac{\partial(ZT)}{\partial T} + \rho\,\frac{\partial^2 (ZT)}{\partial\rho\,\partial T},
\end{equation}
where the first one is the density version of $(\partial C_V/\partial V)_T=T(\partial^2P/\partial T^2)_V$ and the second one follows from $\Gamma = Z+\rho\,\partial Z/\partial\rho$, we obtain
\begin{equation}
a c_{\tilde x} - T\left(\frac{\partial c_{\tilde x}}{\partial T}\right)_{\tilde x} = B_a.
\end{equation}
This relation leads to Eq.~\eqref{e.Lx}, and, together with $(\partial\tilde Y/\partial\tilde x)_\rho$ and $a=1$, to Eqs.~\eqref{e.Dratio} and \eqref{e.Dratio2}; together with Eq.~\eqref{e.Yx}, it leads to Eq.~\eqref{e.DratioU}.

\bibliography{tttLJ.bib}

\end{document}